%% file: main.tex
\documentclass{article}

\usepackage{microtype}
\usepackage{graphicx}
\usepackage{subcaption}
\usepackage{caption}
\usepackage{booktabs} 
\usepackage{float} 
\usepackage{hyperref}
\usepackage{xurl}
\usepackage{enumitem}
\usepackage[export]{adjustbox}

\usepackage[accepted]{mlsys2025}
\makeatletter
\renewcommand{\mlsys@appearing}{}
\makeatother
\mlsystitlerunning{Flex-MIG: Enabling Distributed Execution on MIG}
\usepackage{amsmath}
\usepackage{natbib}
\usepackage[utf8]{inputenc}
\usepackage{kotex}  
\usepackage{graphicx}
\usepackage{verbatim}
\usepackage{enumitem}
\usepackage{booktabs, makecell}
\usepackage[normalem]{ulem}
\begin{document}

\twocolumn[
\mlsystitle{Flex-MIG: Enabling Distributed Execution on MIG}



\mlsyssetsymbol{equal}{*}

\begin{mlsysauthorlist}
\mlsysauthor{Myeongsu Kim}{skku}
\mlsysauthor{Ikjun Yeom}{skku}
\mlsysauthor{Younghoon Kim}{ajou}
\end{mlsysauthorlist}

\mlsysaffiliation{skku}{Department of Computer Science and Engineering, SungKyunKwan Univeristy, Suwon, South Korea}
\mlsysaffiliation{ajou}{Department Software and Computer Engineering, Ajou Univeristy, Suwon, South Korea}
\mlsyscorrespondingauthor{Younghoon Kim}{kyhoon@gmail.com}


\mlsyskeywords{Machine Learning, Multi-Instance GPU, Distrubuted Training, MLSys}

\vskip 0.3in

\begin{abstract}
GPU clusters in multi-tenant settings often suffer from underutilization, making GPU-sharing technologies essential for efficient resource use. 
Among them, NVIDIA Multi-Instance GPU (MIG) has gained traction for providing hardware-level isolation that enables concurrent workloads without interference. 
However, MIG’s hardware rigidity and the conventional one-to-one allocation model jointly lead to severe fragmentation and cluster-wide underutilization. 
We present Flex-MIG, a software-only framework that replaces one-to-one with a one-to-many allocation model and enables host–shared-memory collectives across MIG instances without hardware modification. 
Flex-MIG eliminates drain-required reconfiguration, reduces fragmentation, and improves makespan by up to 17\% across diverse traces, showing that rethinking MIG’s operational model as a software-coordinated layer substantially improves cluster efficiency.
\end{abstract}
]



\printAffiliationsAndNotice{}  
\input{sections/1_Introduction}

\input{sections/2_Background_Motivation}
\input{sections/3_Proposed_Approach}

\input{sections/4_Implementation}

\input{sections/5_Evaluation}
\input{sections/7_Relatedwork}

\input{sections/8_Conclusion}
\bibliography{reference}
\bibliographystyle{mlsys2025}

\appendix
\section{A Supported MIG profiles}
\label{sec:support-mig-profiles}
Table~\ref{tab:mig-profiles} lists the MIG profiles supported on NVIDIA A100-40GB GPUs, 
including the fraction of streaming multiprocessors (SMs), memory size, 
and the maximum number of allocatable instances per GPU.
\begin{table}[h]
    \centering
    \small
    \setlength{\tabcolsep}{4pt}
    \renewcommand{\arraystretch}{1.3}
    \resizebox{\columnwidth}{!}{%
    \begin{tabular}{lccc}
        \toprule
        Profile & Fraction of SMs & Memory & Max allocatable instances per GPU \\
        \midrule
        \textnormal{1g.5gb}  & 1/7 & 5GB  & 7 \\
        \textnormal{1g.10gb} & 1/7 & 10GB & 4 \\
        \textnormal{2g.10gb} & 2/7 & 10GB & 2 \\
        \textnormal{3g.20gb} & 3/7 & 20GB & 2 \\
        \textnormal{4g.20gb} & 4/7 & 20GB & 1 \\
        \textnormal{7g.40gb} & 7/7 & 40GB & 1 \\
        \bottomrule
    \end{tabular}
    }
    \caption{Available MIG profiles on NVIDIA A100-40GB PCIe GPU. 
    Profile \texttt{ig.jgb} denotes $i/7$ of SMs and $j$\,GB of memory; 
    the rightmost column shows the maximum number of such instances per GPU.}
\label{tab:mig-profiles}
\end{table}
\section{Detailed Experimental Setup}\label{sec:Detailed Experimental Setup}
Experiments are conducted on a server equipped with dual Intel Xeon Gold~6338N CPUs, 
two NVIDIA A100~40GB PCIe GPUs (each on a separate NUMA node), 
and a 200~Gbps Mellanox ConnectX-6 NIC. 
We use CUDA~12.2~\cite{cuda-c-programming-guide}, Python~3.8.10, and NCCL~2.21.5~\cite{NCCL}. 
All workloads are executed in a virtualized environment managed by Kubernetes~1.32 (server~1.32.9, client~1.32.5).


\end{document}

%% file: sections/1_Introduction.tex

\section{Introduction}\label{sec:introduction}
The rising demand for AI workloads makes efficient GPU utilization in multi-tenant clusters increasingly critical. In production, small-to-medium models~(e.g., ResNet, MobileNet) are widely deployed due to their practicality. Yet these workloads often underutilize a single GPU due to modest compute and memory needs; prior studies report low device utilization for common training jobs on NVIDIA V100-16GB GPUs \cite{Orion}. This underutilization motivates stronger GPU sharing mechanisms such as NVIDIA’s Multi-Process Service~(MPS)~\cite{nvidia-mps} and Multi-Instance GPU~(MIG)~\cite{nvidia-mig-a100}. MIG is generally better suited to multi-tenant settings because it provides hardware-level isolation~(SMs, L2, memory), reducing interference compared to MPS, and it is supported by major cloud providers \cite{AWS_EKS, RedHat_Openshift, NVIDIA_MIG_K8S}.

Despite MIG’s isolation benefits, our measurements reveal persistent cluster-level inefficiencies arising from the combination of hardware and operational constraints---summarized as the following four constraints~(C1--C4)---and further amplified by the prevailing one-to-one~(single-job-to-single-MIG-instance) allocation model.
\begin{description}[leftmargin=0pt,labelsep=0.5em]
    \item[\textbf{C1 - Fixed profiles}.] MIG only supports predefined, fixed instance profiles~(Table \ref{sec:support-mig-profiles}); arbitrary sizing is impossible.
    \item[\textbf{C2 - Tree-constrained merging}.] Due to MIG’s tree-structured resource layout~\cite{MIGER}, smaller instances often cannot be merged.
    \item[\textbf{C3 - No cross-GPU aggregation}.] MIG instances residing on physically different GPUs cannot be combined for a single job.
    \item[\textbf{C4 - Drain-required reconfiguration}.] Reconfiguring MIG in virtualized clusters typically requires draining all workloads on the target GPU, incurring reconfiguration latency and suspend/resume overheads \cite{nvidia-mig-manager}.
\end{description}
While C1--C3 are hardware-level constraints, C4 is an operational constraint. 
When coupled with the prevailing \emph{one-to-one} allocation model~(M), these constraints collectively result in cluster-level inefficiencies~(I):
\begin{description}[leftmargin=0pt,labelsep=0.5em]
  \item[\textbf{I1 - Over-provisioning}~(\textit{C1 + M}).]
  If a job’s request does not match a predefined profile, it must be allocated to a larger single instance in such one-to-one model, incurring over-provisioning~(Section~\ref{sec:over-provisioning}).
  \item[\textbf{I2 - Internal/External fragmentation}~(\textit{C2 + C3 + M}).]
  Tree-constrained merging and the lack of cross-GPU aggregation prevent composing the right-sized shape under one-to-one allocation, increasing queuing delay~(Section~\ref{sec:resource-fragmentation}).
  \item[\textbf{I3 - Reconfiguration cost}~(\textit{C4, exacerbated by M}).]
  Under the one-to-one allocation model~(M), schedulers more frequently resort to reconfiguration to chase exact-profile fits, increasing the rate of drain-required events. Each drain incurs reconfiguration latency and suspend/resume overheads that are \textbf{orders of magnitude larger than per-request inference latencies}~(Section~\ref{sec:drain-reconfig}). 
\end{description}

To overcome these inefficiencies, we propose \textbf{Flex-MIG}, a software-only framework that enables \textit{one-to-many} allocation model—executing one job across multiple MIG instances. Flex-MIG fixes each GPU to minimum-sized instance and enables NCCL’s Host Shared Memory~(SHM) collectives between MIG instances, thereby logically aggregating MIG instances both within and across GPUs.
This design achieves fine-grained allocation~(mitigating I1), flexible aggregation~(mitigating I2), and reconfiguration-free operation~(mitigating I3), transforming rigid hardware partitioning into a software-managed resource layer. Additionally, realizing one-to-many introduces orchestration considerations~(e.g., determining the instance type and placement) and a runtime challenge~(enabling SHM collectives across MIGs). Flex-MIG addresses these through MIG-aware instance selection and NCCL modifications, as detailed in Section~\ref{sec:proposed}.

This paper makes the following contributions:
\begin{enumerate}
    \item We identify how MIG’s constraints~(C1–C4) and one-to-one allocation model cause inefficiencies~(I1–I3), and propose \textbf{Flex-MIG}, a software-only framework that overcomes them via  a one-to-many allocation model.
    \item We extend NCCL with MIG-aware peer discovery and Host Shared Memory (SHM) collectives, enabling efficient communication across MIG instances on the same host.
    \item In trace-driven experiments, Flex-MIG improves cluster throughput and shortens makespan by up to 17\%, without requiring GPU drain or physical reconfiguration.
\end{enumerate}
The remainder of this paper elaborates on the motivation and technical background behind these observations~(Section~\ref{sec:bg-motivation}), 
before detailing the Flex-MIG design that realizes the proposed one-to-many model.

%% file: sections/2_Background_Motivation.tex
\section{Background \& Motivation}\label{sec:bg-motivation}
\subsection{Scope}\label{sec:scope-assumptions}
\textbf{\textit{Workload Characteristics.}}
We focus on multi-tenant GPU clusters that use NVIDIA MIG for hardware-level isolation and safe concurrency. We target small-to-medium models that benefit from spatial partitioning~\cite{nvidia-mig-a100, nvidia-mig-benefits}, and exclude jobs that already saturate a full GPU or require multi-GPU scaling~(e.g., LLMs, long-running pretraining).

\subsection{Background: GPU Sharing with MIG}\label{sec:gpu-sharing}
\begin{figure}[t] 
    \centering
    \includegraphics[width=\linewidth]{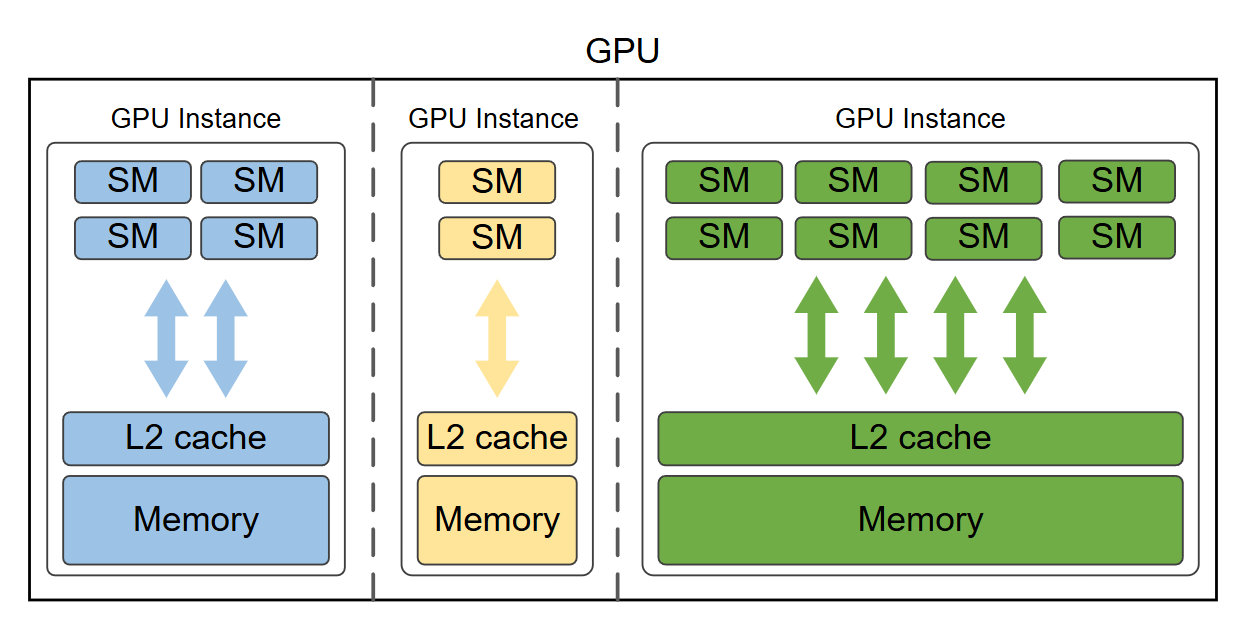}       
    \vspace{-2em}
    \caption{\textbf{MIG architecture}. MIG partitions a GPU into hardware-isolated instances with dedicated SM/L2/memory slices.}
    \label{fig:mig}
    \vspace{1em}
\end{figure}

MIG~(Figure~\ref{fig:mig}), first introduced in the NVIDIA Ampere architecture, partitions a single GPU into multiple hardware-isolated instances, each with dedicated SMs, L2 cache, and memory slices. On an NVIDIA A100-40GB PCIe GPU, up to seven such instances can be created under predefined profiles~(listed in Appendix~\ref{sec:support-mig-profiles}). This hardware-level partitioning provides strong performance and fault isolation, in contrast to MPS, which shares L2 and memory resources and thus suffers from potential interference~\cite{Orion, zhang_migperf_2023}.

\subsection{Inefficiency of the One-to-One Allocation Model}\label{sec:one-to-one inefficiency}
While MIG’s isolation enables safe concurrency, its hardware-level constraints---when coupled with the conventional one-to-one allocation model---amplify inefficiencies at the cluster level, degrading overall utilization and throughput. This subsection summarizes where and why MIG becomes inefficient under this model.

\begin{figure}[t]
    \centering
    \includegraphics[width=\linewidth,height=7cm]{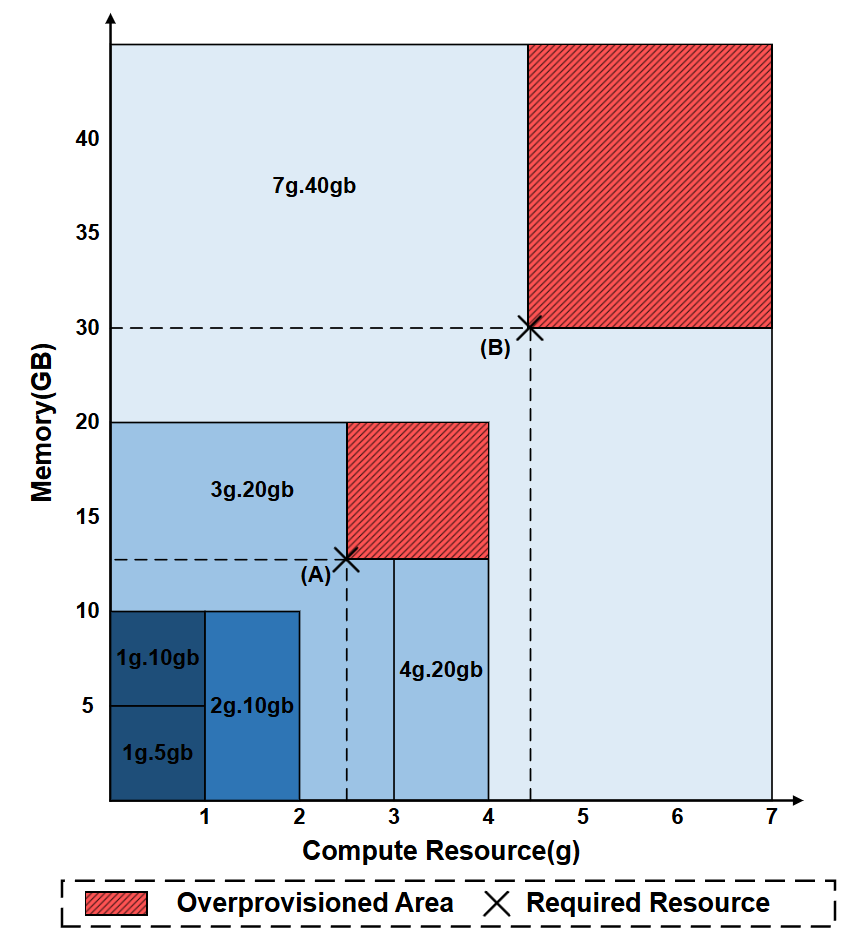}
    \caption{\textbf{Coarse-grained MIG profiles and over-provisioning.} The x-axis and y-axis represent the fractions of total compute units~(SMs) and memory capacity~(GB), respectively. Each rectangle denotes a fixed MIG profile. When a workload’s demand falls between profiles, it is rounded up to the nearest larger profile. The red shaded areas indicate the amount of excess compute or memory allocated in examples~(A) and~(B).}
    \vspace{-2em}
    \label{fig:coarse-grain}
\end{figure}

\subsubsection{Over-provisioning}\label{sec:over-provisioning}
MIG provides a fixed set of instance profiles; arbitrary configurations (e.g., \texttt{3g.15gb}, \texttt{5g.25gb}) are not supported. 
As illustrated in Figure~\ref{fig:coarse-grain}, a 3g/15\,GB or 5g/25\,GB workload is rounded up to \texttt{4g.20gb} or \texttt{7g.40gb}, respectively—wasting compute and memory resources and lowering overall efficiency.

\begin{figure}[t]
    \centering
    \begin{subfigure}[t]{0.9\linewidth}
        \centering
        \includegraphics[width=\linewidth]{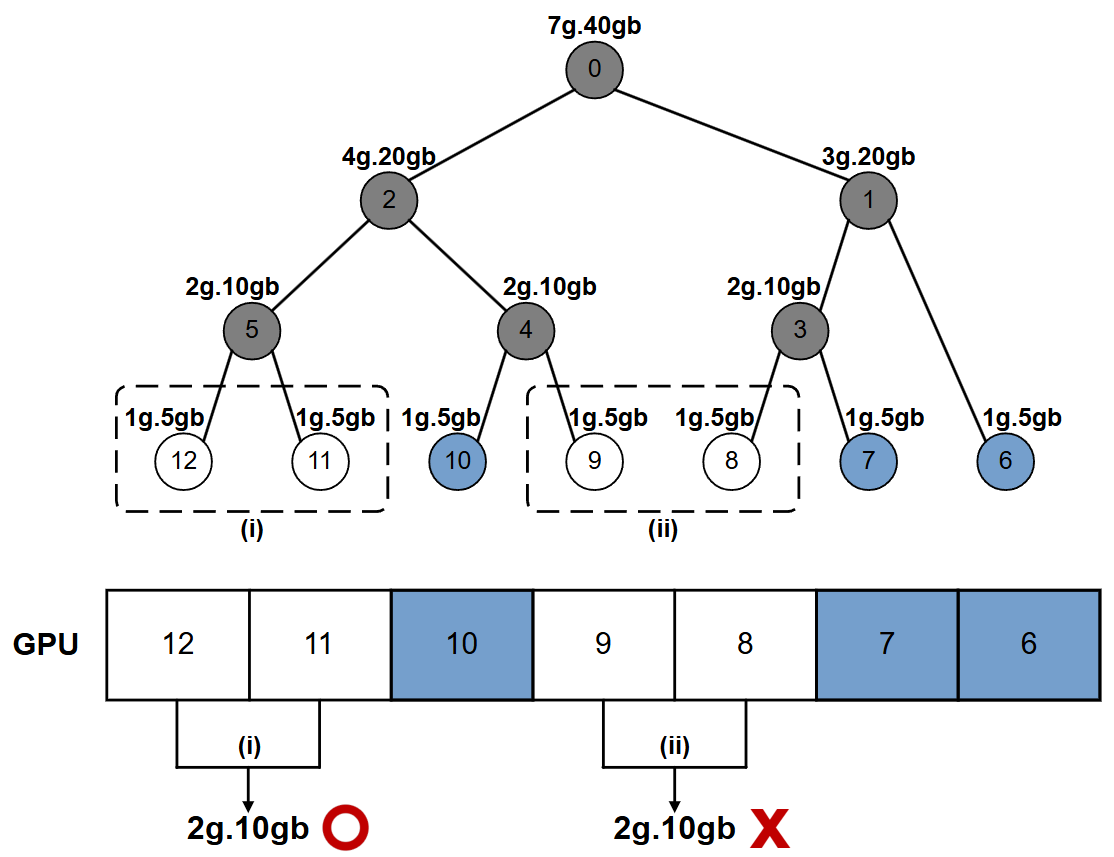}
        \caption{Internal fragmentation}
        \label{fig:internal-frag}
    \end{subfigure}
    \vspace{0.6em}
    \begin{subfigure}[t]{0.9\linewidth}
        \centering
        \includegraphics[width=\linewidth]{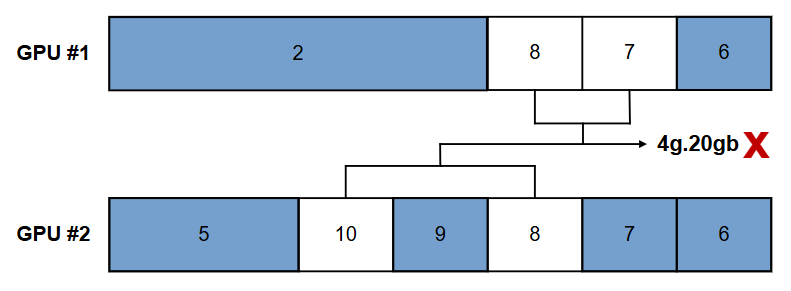}
        \caption{External fragmentation}
        \label{fig:external-frag}
    \end{subfigure}
    \caption{\textbf{Fragmentation under MIG.}
    (a) \emph{Internal fragmentation:} 
    Due to the tree-constrained layout, case~(i) can be merged into \texttt{2g.10gb}; case~(ii) cannot be merged. 
    (b) \emph{External fragmentation:} 
    instances on physically distinct GPUs cannot be merged into a larger instance.}
    \label{fig:mig-fragmentation}
    \vspace{-1.5em}
\end{figure}

\subsubsection{Resource Fragmentation}\label{sec:resource-fragmentation}
MIG exhibits two forms of fragmentation: \emph{internal}~(within a single GPU) and \emph{external}~(across GPUs in a host).

\textbf{\textit{Internal fragmentation.}}
Even when multiple small instances reside contiguously on the same GPU, they may not be mergeable because MIG resources follow a tree layout—adjacency alone is insufficient; only instances sharing the same parent node can be combined into a larger instance~(Figure~\ref{fig:mig-fragmentation}a).

\textbf{\textit{External fragmentation.}}
Instances on different physical GPUs cannot be aggregated to serve a single larger demand~(Figure~\ref{fig:mig-fragmentation}b).

Both cases increase queuing delay and reduce cluster utilization because idle resources, though sufficient in total, cannot be merged into a single larger allocation.

\subsubsection{Drain-required Reconfiguration}\label{sec:drain-reconfig}
Reconfiguring MIG~(merging/splitting instances) in virtualized cluster settings typically \emph{requires draining} all workloads on the target GPU. 

In Kubernetes deployments, NVIDIA's MIG management stack~(e.g., \texttt{mig-manager}) documents that reconfiguration requires the GPU to be idle~\cite{nvidia-mig-manager,NVIDIA_MIG_K8S,nvidia-mig-parted}. Practically, a reconfiguration cycle entails~(i) suspending running jobs and saving state~(checkpoint), (ii) executing the MIG reconfigure operation, and (iii) restarting jobs and recreating pods, each incurring non-trivial overheads.

We measured some of these overheads on our testbed using the target models in Table~\ref{tab:workload-list}. Because our models are small-to-medium scale, each checkpoint \textit{save} and \textit{load} operation typically incurred a few seconds of overhead. By contrast, the \textit{MIG reconfigure} operation driven by \texttt{mig-manager} took 100--120\,s end to end in our environment~(from reconfigure request to the system reporting a stable, ready state). Pod \textit{delete/create} added additional seconds. These measurements align with prior reports that reconfiguration commonly incurs seconds to tens of seconds of latency~\cite{MIGER,MISO}.

\subsection{Rationale for One-to-Many}\label{sec:rationale one-to-many}
We posit that replacing the conventional one-to-one allocation model with \emph{one-to-many} mitigates cluster-level inefficiencies under MIG.
In one-to-many, GPUs are statically partitioned into the smallest leaves~(e.g., \texttt{1g.5gb}×6 + \texttt{1g.10gb}×1 on an A100-40GB PCIe), and a single workload is scheduled across multiple MIG instances in a distributed manner. 
This achieves:
\begin{enumerate}
    \item \textbf{Resource flattening.} 
    Fixing all GPUs to the smallest profile creates a uniform pool of interchangeable leaves.  
    Demands that do not exist as a single profile~(e.g., \texttt{5g.25gb}, \texttt{6g.30gb}) can be satisfied by assigning 5 or 6 \texttt{1g.5gb} instances to the same job \textit{(resolves I1)}. 
    Moreover, leaves that were previously unmergeable within a GPU due to the tree constraint~(Figure~\ref{fig:internal-frag}) or distributed across different GPUs~(Figure~\ref{fig:external-frag}) can be \emph{logically aggregated} by the orchestration layer, reducing both internal and external fragmentation \textit{(resolves I2)}.
    \item \textbf{Reconfiguration avoidance.} 
    Fine-grained logical aggregation across leaves eliminates the need for physical reconfiguration, avoiding GPU drain and its associated overheads \textit{(resolves I3)}.
\end{enumerate}
In summary, logical aggregation across multiple MIG instances increases scheduling flexibility and enables continuous workload execution without disruptive reconfiguration.

\subsection{Challenges for One-to-Many}\label{sec:challenges for one-to-many}
We focus here on the runtime challenge of one-to-many under MIG; design-side orchestration considerations are handled by the orchestration layer and discussed later.

Distributed deep learning relies on collective communication among participating devices. NVIDIA NCCL is the de facto backend for GPU clusters~\cite{pytorch_DDP_NCCL,NCCL}. NCCL forms a communicator among ranks and selects transport paths~(e.g., NVLink, PCIe P2P, Host Shared Memory-SHM, or network RDMA/IP) based on the discovered hardware topology. Across-MIG NVLink and PCIe P2P are officially unsupported~\cite{nvidia-mig-a100}; therefore, for ranks on the same host but in different MIG instances, efficient execution hinges on using SHM rather than network paths.

However, when multiple MIG instances from the same physical GPU join one communicator, NCCL's PCIe Bus ID–based device identification causes false duplicate detection, since all instances on the same GPU share an identical Bus ID.
We observe two failure points:
\begin{itemize}
    \item \textbf{Peer discovery.}
    During rank information exchange---rank IDs, device IDs, host hashes, PID hashes---NCCL performs duplicate-GPU checks. If two ranks bind MIG instances that share a Bus ID, NCCL misclassifies them as the same device and aborts.
    \item \textbf{Topology construction.}
    While building the system topology, NCCL registers devices incrementally. If a newly added device has a Bus ID already present, NCCL deduplicates it—mistakenly collapsing distinct MIG instances into one node. The resulting topology has fewer devices than ranks, causing communicator construction to fail.
\end{itemize}
A naive workaround isolates instances into separate containers/VMs so they appear as different hosts, which forces network transports~(RDMA/IP) and hurts efficiency. 
Therefore, for the one-to-many operational model to work efficiently, effective collective communication across MIG instances is essential.
This, in turn, requires a mechanism within NCCL to distinguish individual MIG instances during peer discovery and topology construction.

These characteristics and limitations motivate a new allocation model that can flexibly combine multiple MIG instances for a single job — the rationale for our one-to-many approach, described next.

%% file: sections/3_Proposed_Approach.tex
\section{Proposed Approach}\label{sec:proposed}
We propose \textbf{Flex-MIG}, a software-only framework that enables a new operational model---\emph{one-to-many allocation}---so that a single job can run across multiple MIG instances.
This improves cluster-level efficiency by reducing queuing delay and increasing throughput, without requiring physical reconfiguration.
To fully utilize the GPU memory capacity, we configure each NVIDIA A100-40GB PCIe GPU as six \texttt{1g.5gb} instances and one \texttt{1g.10gb} instance, instead of seven \texttt{1g.5gb} instances, which would leave 5\,GB of memory unused.
This section describes how Flex-MIG differs from the conventional one-to-one operation model and details its core design components.

\subsection{System Overview}\label{sec:system-overview}
\begin{figure*}[t]
\centering
\includegraphics[width=\textwidth]{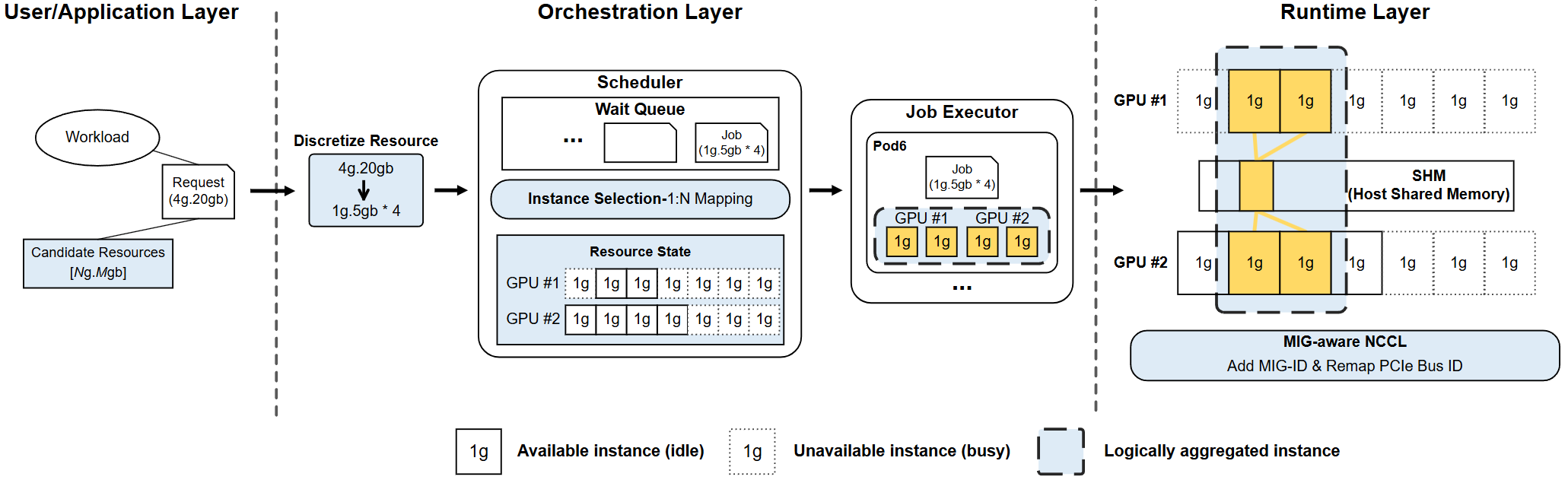}
\caption{\textbf{Flex-MIG system overview.}
Flex-MIG operates over a three-layer cluster abstraction---\emph{user/application}, \emph{orchestration}, and \emph{runtime}---to realize one-to-many execution using fixed, minimum-sized MIG leaves.
The orchestration layer manages job queuing and instance allocation, while the runtime layer enables Host Shared Memory~(SHM) collectives across MIG instances.}
\label{fig:overview}
\end{figure*}

Flex-MIG builds on a three-layer cluster abstraction commonly used in multi-tenant GPU environments: the \emph{user/application}, \emph{orchestration}, and \emph{runtime} layers.
Under this abstraction, Flex-MIG primarily extends the orchestration and runtime layers to realize the one-to-many allocation model.
Figure~\ref{fig:overview} illustrates how the proposed system coordinates and executes one-to-many allocation across these layers.

\textbf{\textit{User/Application Layer.}}
This layer exposes available resource types to users and receives workload requests.
Unlike the baseline MIG environment, Flex-MIG allows finer-grained compute and memory requirements (e.g., \texttt{5g.15gb}, \texttt{6g.30gb}) independent of fixed profiles. Although Flex-MIG cannot always achieve a perfect fit due to the inherent 1g–5GB coupling in hardware, finer-grained rounding reduces over-provisioning and leaves more residual capacity for other jobs.

\textbf{\textit{Orchestration Layer.}}
This layer manages job queuing, admission, and instance selection across available resources in a virtualized cluster environment. 
In Flex-MIG, its primary responsibility is to determine \emph{which} MIG instances to allocate and \emph{how} to place them for a single job under the one-to-many model.
Detailed mechanisms are discussed in Section~\ref{sec:orchestration-layer}.

\textbf{\textit{Runtime Layer.}}
This layer handles process creation and environment setup for actual job execution.
Flex-MIG fixes each GPU to minimum-sized instances and \emph{logically aggregates} multiple instances to run a single job in a distributed manner. Efficient collective communication across MIG instances is therefore required~(Section~\ref{sec:runtime-layer}).

\subsection{Orchestration Layer}\label{sec:orchestration-layer}
The orchestration layer in Flex-MIG consists of a \textbf{scheduler}, which queues and assigns MIG instances to incoming jobs, and a \textbf{job executor}, which launches and manages Pods for scheduled workloads and reclaims their resources upon completion.

Transitioning to Flex-MIG’s \emph{one-to-many} from one-to-one allocation introduces new design considerations for scheduling.
Specifically, the scheduler must determine:
(i) \emph{which types of instances} to allocate among heterogeneous instance profiles (e.g., \texttt{1g.10gb} vs.\ \texttt{1g.5gb}), and 
(ii) \emph{where} to place those instances across physical GPUs (on the same GPU or distributed across GPUs within a host). 

To address these, Flex-MIG introduces an \textbf{instance selection policy} composed of two heuristics:
(1) \textbf{size-aware instance prioritization}, and 
(2) \textbf{topology-aware placement}.

\textbf{\textit{Size-Aware Instance Prioritization.}}
We empirically observed that the performance impact of instance type varies depending on workload size.
For workloads defined in Table~\ref{tab:workload-list}, when the workload size is~1, running on \texttt{1g.10gb} improves job completion time (JCT) by 10--30\% compared to \texttt{1g.5gb}.
However, for workload sizes~$\ge$~2, combining \texttt{1g.10gb} and \texttt{1g.5gb} instances yields no noticeable benefit, 
because distributed training across multiple MIG instances involves synchronization overhead, effectively limiting performance to that of the slower \texttt{1g.5gb} instances.
Based on these observations, Flex-MIG applies a simple heuristic that prioritizes \texttt{1g.10gb} for single-instance jobs (size~=~1) and \texttt{1g.5gb} for multi-instance jobs (size~$\ge$~2).

\textbf{\textit{Topology-Aware Placement.}}
When scheduling jobs that require multiple instances, the scheduler must also determine \textit{where} to place them:
(i) allocating multiple instances on the same GPU, or
(ii) distributing instances across different GPUs within a host.
Our experiments revealed that distributing instances evenly across physical GPUs yields the best performance (Section~\ref{sec:job_perf}).
Consequently, Flex-MIG adopts a \textit{round-robin placement policy} that allocates MIG instances evenly across GPUs within a node to maximize overall throughput.

\subsection{Runtime Layer}\label{sec:runtime-layer}
In the conventional one-to-one model, communication between MIG instances does not occur; however, Flex-MIG’s one-to-many model requires collective communication across multiple instances.
As discussed in Section~\ref{sec:challenges for one-to-many}, the current NCCL GPU identification logic fails to distinguish different MIG instances on the same GPU.

To address this issue, we embed a unique \texttt{mig\_id} \textit{identifier} for each instance during peer discovery and assign \textit{synthetic labels} to duplicate PCIe Bus IDs. 
These modifications enable NCCL to correctly identify multiple MIG instances residing on the same GPU and establish SHM-based collective communication across them. 
As a result, Flex-MIG achieves intra-host collectives without relying on network transports or requiring physical reconfiguration, thereby closing the key runtime gap that previously prevented one-to-many operation.

%% file: sections/4_Implementation.tex
\section{Implementation}\label{sec:implementation}
We implement Flex-MIG with two core components: (i) an orchestration layer that operates on Kubernetes~\cite{Kubernetes} via the Python's kubernetes client API~\cite{kubernetes_client_python}, and (ii) a runtime layer built upon NCCL~\cite{NCCL} with minimal source-level modifications. 
Distributed execution is managed through PyTorch’s Distributed Data Parallel~(DDP)~\cite{pytorch_DDP_NCCL} framework.

\subsection{Orchestration Layer}\label{sec:Orchestration-layer-impl}
\subsubsection{Scheduler}\label{sec:Scheduler-impl}
All submitted jobs reside in a global \textit{wait queue}. 
The scheduler repeatedly dequeues jobs and checks resource availability through the Kubernetes API. 
If a job is admissible, it selects concrete MIG instances according to the \textit{instance selection policy}~(Section~\ref{sec:orchestration-layer}) and passes a job descriptor with metadata to the \textit{Job Executor}.

\subsubsection{Job Executor}\label{sec:Job Executor-impl}
The \textbf{Job Executor} bridges scheduling decisions made by the orchestration layer and the actual runtime execution. 
It prepares a pod
    \footnote{A \textit{pod} is the basic execution unit in Kubernetes~\cite{Kubernetes}, encapsulating one or more containers sharing the same resources.}
specification in which the environment variable \texttt{NVIDIA\_VISIBLE\_DEVICES}~\cite{nvidia_container_toolkit_docker} lists the assigned MIG UUIDs, restricting the container’s visibility to those instances.
Each MIG UUID is a globally unique identifier that distinguishes individual MIG instances. 
The Job Executor then launches the pod through the Kubernetes API with an appropriate entrypoint for distributed execution.
\begin{figure}[t]
    \centering
    \includegraphics[width=\linewidth]{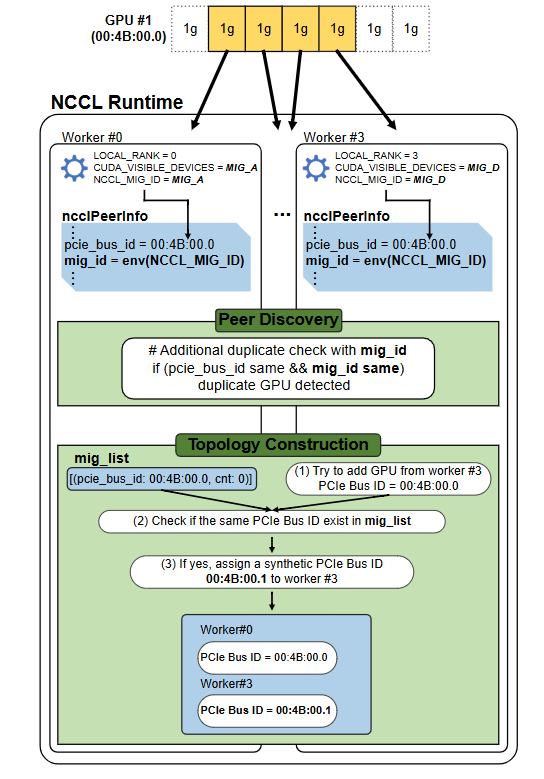}
    \caption{MIG-aware NCCL runtime workflow.}
    \label{fig:nccl_impl}
    \vspace{-1em}
\end{figure}

\subsection{Runtime Layer}\label{sec:Job Runtime-layer-impl}
The runtime layer handles process-level initialization and MIG-aware collective communication. 
To enable SHM-based collectives among different MIG instances on the same GPU, each process must propagate additional binding information to NCCL.

During job launch, multiple worker processes are spawned—one per assigned MIG instance—and each is identified by its \texttt{LOCAL\_RANK}.
Each process reads the pod-level environment variable \texttt{NVIDIA\_VISIBLE\_DEVICES} to obtain its assigned MIG UUID, 
then exports it to both \texttt{CUDA\_VISIBLE\_DEVICES} (for CUDA binding)~\cite{nvidia_deploy_topic_5_2_1} and \texttt{NCCL\_MIG\_ID} (for NCCL identification).
This per-process initialization occurs within the container runtime and ensures that NCCL receives the correct instance-level metadata before communicator setup.

We modify two parts of NCCL~(Figure~\ref{fig:nccl_impl}) to enable SHM-based collectives among different MIG instances on the same GPU: 
(i) \textit{MIG-aware Peer Discovery} and (ii) \textit{Synthetic Bus-ID Labeling}. 
We do not alter any collective algorithms or transport primitives.

\subsubsection{MIG-aware Peer Discovery}\label{sec:peer-discovery-impl}
We introduce a new field, \texttt{mig\_id}, in the peer metadata structure exchanged among ranks. As shown in the top of Figure~\ref{fig:nccl_impl}, each worker assigns the value of \texttt{NCCL\_MIG\_ID} to \texttt{mig\_id}. We then modify the duplicate-checking logic to additionally compare \texttt{mig\_id}~(Peer Discovery in Figure~\ref{fig:nccl_impl}). Thus, even when two peers report the same \texttt{pcie\_bus\_id}, they are treated as distinct devices if their \texttt{mig\_id} differs. Because \texttt{mig\_id} carries the actual MIG UUID, accidental double-binding of the \emph{same} instance is still correctly detected.

\subsubsection{Synthetic Bus-ID Labeling}\label{sec:Synthetic Bus-ID Labeling-impl}
We introduce a \emph{synthetic Bus-ID labeling} mechanism~(Topology Construction part in Figure~\ref{fig:nccl_impl}). 
During topology construction, we maintain a list of \texttt{(pcie\_bus\_id, count)} tuples, which we call \texttt{mig\_list}. 
When a new rank references a Bus ID already present in \texttt{mig\_list}, we increment the counter and append a synthetic suffix to that Bus ID~(e.g., \texttt{00:4B:00.0} $\rightarrow$ \texttt{00:4B:00.1}). 
This ensures that each rank is registered as a unique topology node without altering NCCL’s APIs. 
Because these suffixed Bus IDs are not valid device identifiers, any path that would pass them to a driver is guarded by a restoration routine that strips the synthetic suffix before use.

These mechanisms enable SHM-based collectives among distinct MIG instances on the same host, making the one-to-many execution model operationally practical.

%% file: sections/5_Evaluation.tex
\section{Evaluation}\label{sec:evaluation}
The goal of this section is to verify whether Flex-MIG improves cluster-wide resource utilization and throughput compared to existing MIG operation models. However, conventional MIG requires GPU drain upon reconfiguration, which makes repeated experiments across diverse workloads and system states extremely difficult. Moreover, executing multi-condition experiments with many concurrent workloads on a real cluster is impractical. Therefore, we build a calibrated simulator based on real Flex-MIG executions and use it to compare against other MIG operation modes.

We answer the following questions:
\begin{enumerate}[itemsep=0.5pt, topsep=0pt, parsep=0pt, partopsep=0pt]
    \item How closely does our simulator reproduce real-world results?
    \item Does Flex-MIG improve cluster throughput and makespan?
    \item How much performance difference exists between one-to-one and one-to-many execution for individual jobs?
    \item Does SHM-based collective communication improve the performance compared to RDMA-based transport?  
\end{enumerate}

\begin{table}[t]
\vskip 0.1in
\centering
\small
\renewcommand{\arraystretch}{1.15}
\resizebox{0.95\columnwidth}{!}{%
\begin{tabular}{lcccc}
\toprule
Model & \multicolumn{2}{c}{Batch size} & \multicolumn{2}{c}{Workload size} \\
\cmidrule(lr){2-3}\cmidrule(lr){4-5}
 & Train & Inference & Train & Inference \\
\midrule
ResNet-18           & 128         & 32         & 1     & 1 \\
ResNet-34           & 256         & 64         & 2     & 2 \\
ResNet-50           & 196,256     & 64         & 4,6   & 4 \\
ResNet-101          & 256         & –          & 8     & – \\
MobileNetV3-Small   & 256–512     & 64–128     & 1–2   & 1–2 \\
MobileNetV3-Large   & 64–512      & 32–128     & 1–6   & 1–4 \\
EfficientNet-B0     & 32–256      & 16–64      & 1–6   & 1–4 \\
EfficientNet-B2     & 32–196,256      & 8–32       & 1–6,8   & 1–4 \\
DistilBERT          & 8–64        & 4–16       & 1–6   & 1–4 \\
BERT-Base           & 4–32        & 2–8        & 1–6   & 1–4 \\
T5-Small            & 16–128      & 8–32       & 1–8   & 1–4 \\
\bottomrule
\end{tabular}%
}
\vspace{-0.4em}
\caption{
Workloads used in evaluation.
Each model is trained and evaluated with representative batch sizes,
expressed as ranges when multiple configurations are tested.
Unless otherwise noted, batch sizes and workload size increase by powers of two.
Models follow standard implementations from the original papers~\cite{resnet,mobilenet,efficientnet,distil-bert,bert,t5},
using publicly available pretrained checkpoints from PyTorch~\cite{Pytorch-pre-training} and Hugging Face~\cite{HF_BERT_base_cased,HF_DistilBERT_base_uncased, HF_T5_small}.
}
\label{tab:workload-list}
\vspace{-0.2em}
\end{table} 

\begin{table}[t]
    \vskip 0.1in
    \begin{center}
        \small
        \resizebox{\columnwidth}{!}{%
        \begin{tabular}{lcc}
        \toprule
        Size category & Train~(1 / 2 / 4 / 6 / 8) & Inference~(1 / 2 / 4) \\
        \midrule
        Small-Dominant    & 16 / 8 / 4 / 2 / 1 & 16 / 8 / 4 \\
        Balanced &  8 / 8 / 8 / 4 / 4 & 10 / 10 / 10 \\
        Large-Dominant    &  4 / 4 / 12 / 8 / 4 & 8 / 8 / 16 \\
        \bottomrule
        \end{tabular}
        }
    \end{center}
   \caption{Workload size distribution across training and inference workloads. 
    Numbers in parentheses~(e.g., 1 / 2 / 4 / 6 / 8 for training and 1 / 2 / 4 for inference) denote workload sizes, 
    and the values in each cell represent the number of jobs for each size and type category.}
    \vspace{-0.4em}
    \label{tab:workload-size-dist}
    \vspace{-0.6em}
\end{table} 

\subsection{Setup}\label{sec:setup}
\textbf{Environment.}
Experiments are conducted on a server with two NVIDIA A100~40GB PCIe GPUs~(details listed in Appendix~\ref{sec:Detailed Experimental Setup}).

Training jobs use PyTorch DDP \cite{pytorch_DDP_NCCL} with ZeRO \cite{pytorch_zero} to reduce per-rank memory footprint, while inference jobs use DDP with an additional \texttt{allgather} operation to aggregate results. 
All traces are executed in an open-loop manner to fairly measure cluster-level metrics such as makespan, average waiting time, and utilization.

\textbf{\textit{Target workloads.}}
Table~\ref{tab:workload-list} summarizes training and inference models and their batch sizes. For the baseline MIG~(one-to-one allocation), workload sizes 2 and 4 correspond to \texttt{2g.10gb} and \texttt{4g.20gb} respectively, while sizes 6–8 require a full \texttt{7g.40gb} instance.

\textbf{\textit{Synthetic traces.}}  
We construct synthetic traces to capture diverse runtime characteristics by combining three orthogonal dimensions:  
(i) execution time distribution,  
(ii) workload size distribution, and  
(iii) workload type~(training or inference).  
Execution time distributions are derived from four public cluster traces—Helios~(Earth, Venus)~\cite{helios-trace}, Philly~\cite{philly-trace}, and Alibaba~\cite{alibaba-trace}.  
We extract single-GPU jobs (or 0.5–1 GPU in the Alibaba trace) and categorize their durations into {short}~(600–1{,}800~s), {medium}~(1{,}800–3{,}600~s), and {long}~(3{,}600–7{,}200~s), following the empirical distribution observed in the original traces.

Workload size distributions are defined as {small-dominant}, {balanced}, or {large-dominant} and workload types are configured as training-only, inference-only, or 50:50 mixed.  
We assume independence among these dimensions and generate all possible combinations to produce a diverse set of synthetic traces.

\textbf{\textit{Scenarios and baselines.}}  
We evaluate Flex-MIG~(FM) against two representative baselines: \textit{Dynamic-MIG~(DM)} and \textit{Static-MIG~(SM)}.  
Dynamic-MIG dynamically reconfigures GPU instances according to job requests.  
When merging is blocked by the tree constraint, it performs GPU drain and reallocation. For inference jobs, GPU drain is prohibited to prevent service interruption.  
Static-MIG uses fixed instance combinations of [\texttt{1g.10gb}, \texttt{2g.10gb}, \texttt{4g.20gb}] and therefore supports workloads up to size~4.  
For Static-MIG, we adopt an additional scheduling rule inspired by~\cite{MIG-Adapter}: if the requested instance type is unavailable, a larger idle instance may be allocated to the job. This adjustment is made to maximize overall throughput under the static partitioning constraint. 

We employ two scheduling policies: \textit{FIFO} and \textit{Aggressive Backfilling}.  
While FIFO scheduling considers only a single candidate job at a time, aggressive backfilling examines multiple candidates (14 in our configuration).
We include Aggressive Backfilling not as a design component of Flex-MIG, but to ensure experimental fairness—specifically, to give Dynamic-MIG and Static-MIG additional opportunities for job placement and throughput improvement.  
Accordingly, we evaluate two scenario groups:  
(i) {Training-only scenarios}: FM vs.\ DM vs.\ SM under FIFO scheduling, and  
(ii) {Mixed~(training+inference) scenarios}: FM vs.\ DM under Aggressive Backfilling.

\textbf{\textit{Metrics.}}  
The experiment using simulator, we double the number of workloads from Table~\ref{tab:workload-size-dist} and generate ten distinct traces per category to ensure statistical diversity and robustness.  
We evaluate following five primary metrics. 
\textbf{Makespan} denotes the total wall-clock time to complete all jobs in a given trace.  
\textbf{Average JCT} represents the mean completion time of individual jobs.  
\textbf{Average Waiting Time} measures the delay between a job becoming eligible for scheduling and its actual start of execution.  
\textbf{Average External Fragmentation Delay}, defined as the cumulative waiting time of jobs that could not be scheduled even when sufficient total GPU resources were available, due to their distribution across different physical GPUs.
\textbf{Cluster Utilization} quantifies the proportion of time during which MIG instances were actively executing workloads between the start of the first job and the completion of the last job.

\begin{figure}[t]
    \centering
        \begin{subfigure}[t]{0.48\linewidth}
            \includegraphics[width=\linewidth]{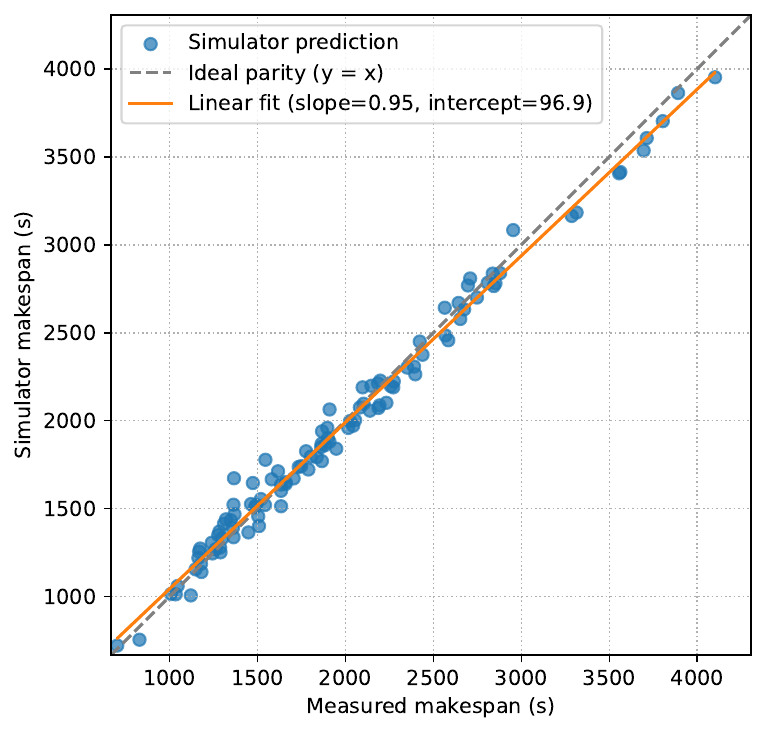}
            \subcaption{Makespan comparison}    
        \end{subfigure}
    \hfill
        \begin{subfigure}[t]{0.48\linewidth}
            \includegraphics[width=\linewidth]{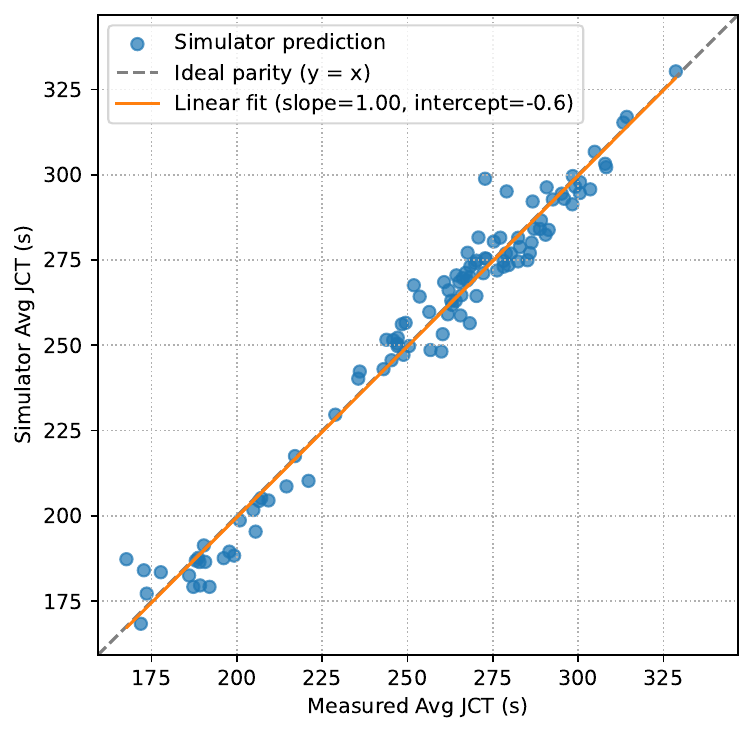}
            \subcaption{Average JCT comparison}    
        \end{subfigure}
    \caption{Parity plots comparing measured and simulated results.
    Each point represents a simulator prediction for a specific scenario. 
    The closer the points lie to the ideal parity line~($y=x$), the more accurate the simulator prediction.}
    \vspace{-1em}
    \label{parity}
\end{figure}

\subsection{Validation of Simulator Accuracy}\label{sec:validation_of_simulator}
We first validate how accurately our simulator reproduces Flex-MIG’s behavior using 108 synthetic traces~(3 traces per category). 
Before building the simulator, we measured the job completion time~(JCT) of individual workloads in real-world environment, taking into account the physical placement of each MIG instance. 
The simulator uses the measured JCTs together with the same scheduling logic as the real Flex-MIG system to emulate its execution behavior. 
Before calibration, the simulator underestimated makespan by approximately 14\% and average JCT by about 6\% compared to real measurements. 
This discrepancy arises because, in the real environment, multiple workloads run concurrently and experience mild performance interference that is not captured in the unadjusted model. 
To compensate for this effect, we applied a constant scaling factor of 1.06 to each workload’s execution time, reflecting the mild resource contention observed in the real cluster.
After calibration, the simulator closely matched the measured results, as shown in Figure~\ref{parity}. 
We use this calibrated simulator for all subsequent comparisons between Flex-MIG and other MIG operation modes.

\begin{figure}[t]
    \centering
    \includegraphics[width=0.6\linewidth, trim=0 30 0 25, clip]{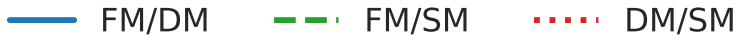}
    \begin{minipage}[t]{0.49\linewidth}
        \centering
        \vspace{0.3em}
        \includegraphics[width=\linewidth]{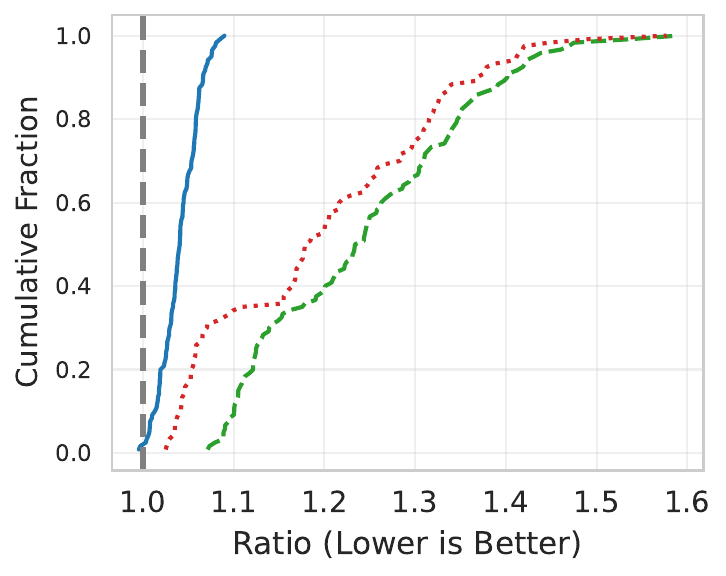}
        \vspace{-1.7em}
        \subcaption{Average JCT}
        \label{fig:fm_dm_sm_cdf_jct}
    \end{minipage}
    \hfill
    \begin{minipage}[t]{0.49\linewidth}
        \centering
        \vspace{0.3em}
        \includegraphics[width=\linewidth]{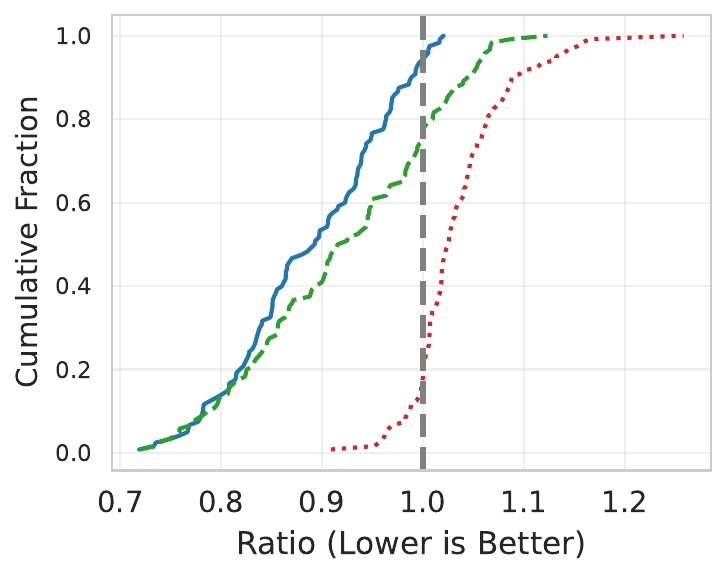}
        \vspace{-1.7em}
        \subcaption{Average Waiting Time}
        \label{fig:fm_dm_sm_cdf_wait}
    \end{minipage}
    \vspace{0.5em}
    
    \begin{minipage}[t]{0.49\linewidth}
        \centering
        \vspace{0.2em}
        \includegraphics[width=\linewidth]{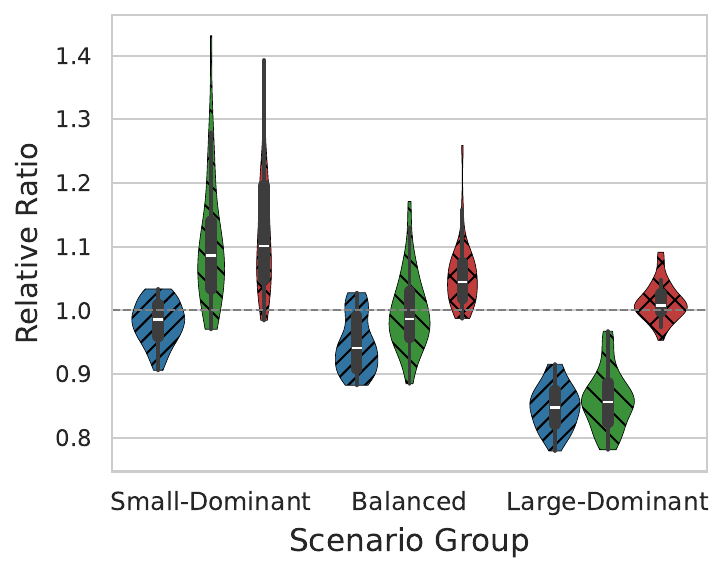}
        \vspace{-1.7em}
        \subcaption{Makespan}
        \label{fig:fm_dm_sm_violin_makespan}
    \end{minipage}
    \hfill
    \begin{minipage}[t]{0.49\linewidth}
        \centering
        \vspace{0.2em}
        \includegraphics[width=\linewidth]{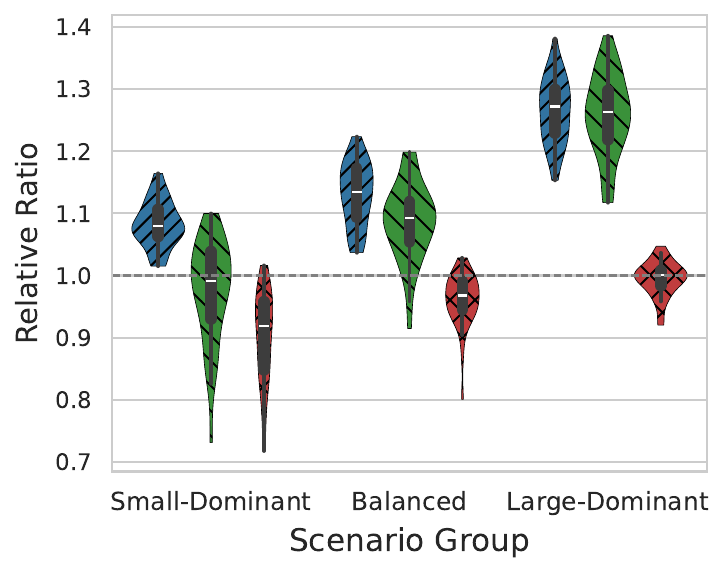}
        \vspace{-1.7em}
        \subcaption{Cluster Utilization}
        \label{fig:fm_dm_sm_violin_util}
    \end{minipage}
    \vspace{0.4em}
    \includegraphics[width=0.6\linewidth, trim=0 30 0 25, clip]{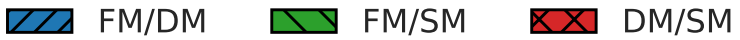}
    \caption{
    Comparison of Flex-MIG~(FM), Dynamic-MIG~(DM), and Static-MIG~(SM). A ratio below~1.0 indicates that the numerator system~(e.g., FM) achieves a lower metric value—i.e., lower JCT, waiting time, or makespan—than the denominator.
    }
    \label{fig:fm_dm_sm}
    \vspace{-1em}
\end{figure}

\begin{figure}[t]
     \centering
     \begin{minipage}[t]{0.5\linewidth}
        \centering
         \includegraphics[width=0.7\linewidth, trim=0 30 0 25, clip]{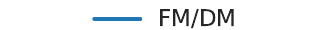}
        \vspace{-2em}
    \end{minipage}
    \hfil
    \begin{minipage}[t]{0.49\linewidth}
        \centering
        \includegraphics[width=\linewidth]{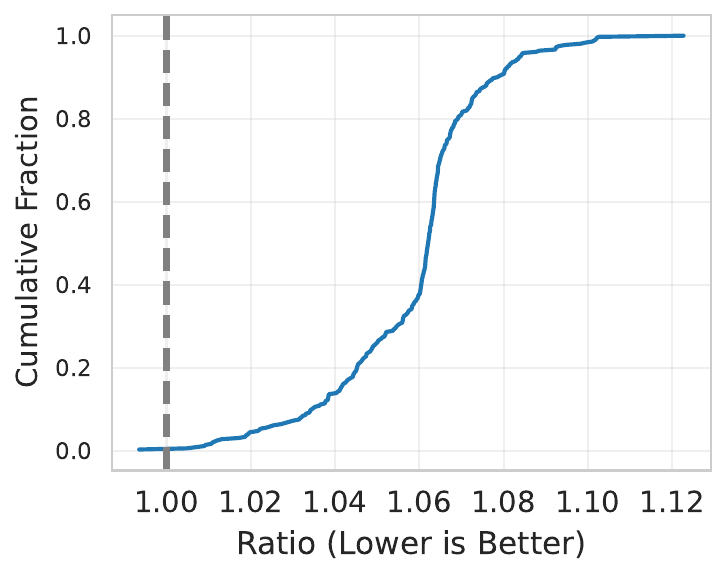}
        \vspace{-1.7em}
        \subcaption{Average JCT}
        \label{fig:fm_dm_cdf_jct}
    \end{minipage}
    \hfill
    \begin{minipage}[t]{0.49\linewidth}
        \centering

        \includegraphics[width=\linewidth]{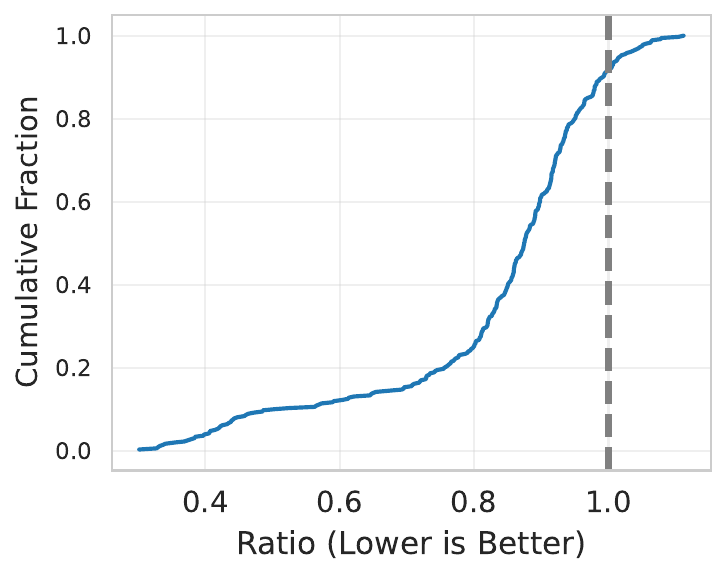}
        \vspace{-1.7em}
        \subcaption{Average Waiting Time}
        \label{fig:fm_dm_cdf_wait}
    \end{minipage}
    

    \vspace{0.5em}
    
    \begin{minipage}[t]{0.49\linewidth}
        \centering
        \vspace{0.2em}
        \includegraphics[width=\linewidth]{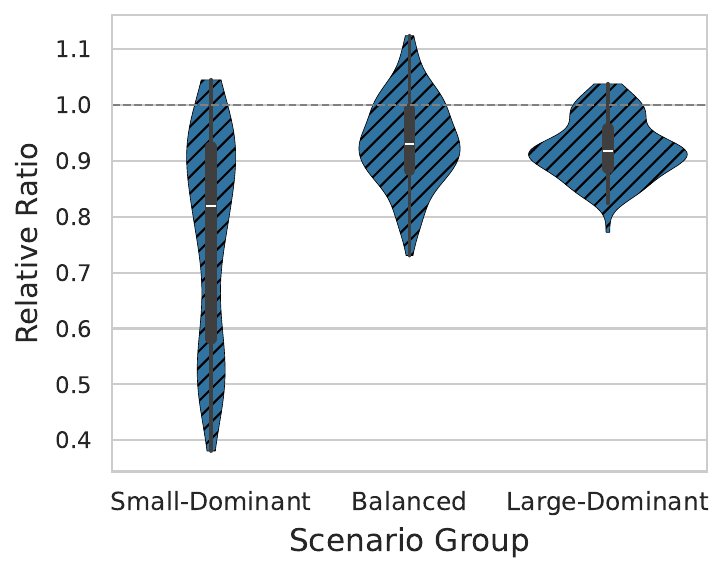}
        \vspace{-1.7em}
        \subcaption{Makespan}
        \label{fig:fm_dm_violin_makespan}
    \end{minipage}
    \hfill
    \begin{minipage}[t]{0.49\linewidth}
        \centering
        \vspace{0.2em}
        \includegraphics[width=\linewidth]{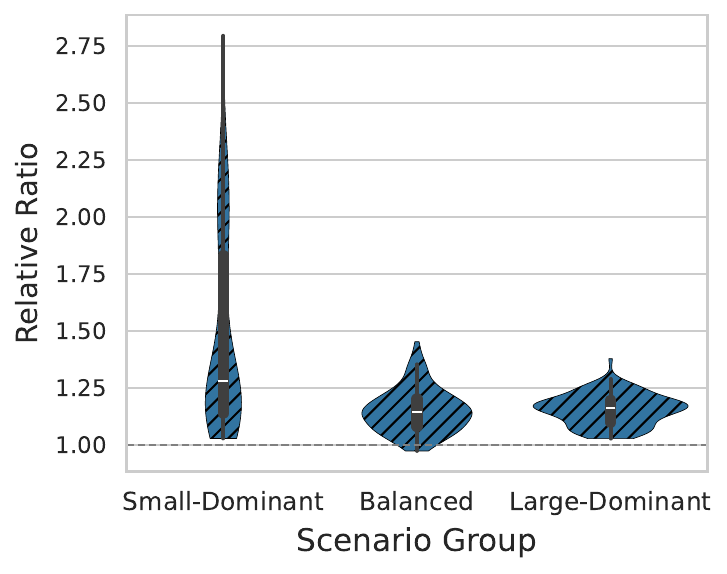}
        \vspace{-1.7em}
        \subcaption{Cluster Utilization}
        \label{fig:fm_dm_violin_util}
    \end{minipage}

    \vspace{0.4em}
    \includegraphics[width=0.4\linewidth, trim=0 30 0 25, clip]{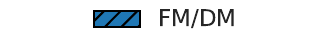}
    \caption{Comparison of Flex-MIG~(FM), Dynamic-MIG~(DM).}
    \vspace{-1.5em}
    \label{fig:fm_dm}
\end{figure}

\subsection{Effectiveness of Flex-MIG}\label{sec:effectiveness}
Figure~\ref{fig:fm_dm_sm} compares Flex-MIG~(FM), Dynamic-MIG~(DM), and Static-MIG~(SM) under \textsc{FIFO} for training-only scenarios with a maximum workload size of~4.

Figures~\ref{fig:fm_dm_sm_cdf_jct} and~\ref{fig:fm_dm_sm_cdf_wait} show the distributions of average JCT and average waiting time across the three operation modes.
Static-MIG attains the lowest per-job JCT because, when the requested instance is unavailable, it can allocate a larger idle instance.
However, Static-MIG admits at most three concurrent workloads per GPU (due to fixed partitioning), which lengthens queues despite its short per-job JCT.
Dynamic-MIG reduces waiting time relative to Static-MIG by more closely fitting jobs via on-demand reconfiguration. Flex-MIG, by contrast, substantially lowers average waiting time---11\% lower than Dynamic-MIG on average---while incurring only a modest increase in per-job JCT (about 4\% on average and never exceeding {10\%).

Figures~\ref{fig:fm_dm_sm_violin_makespan} and~\ref{fig:fm_dm_sm_violin_util} summarize makespan and cluster utilization.
Flex-MIG consistently achieves the shortest makespan and the highest utilization among all modes.
In {large-dominant} traces, Flex-MIG reduces makespan by roughly 15\% relative to Dynamic-MIG.
The gap stems from Dynamic-MIG’s inability to aggregate fragmented resources across GPUs: large jobs must wait until a single GPU exposes a sufficiently large instance.
In contrast, Flex-MIG’s logical aggregation across MIG instances allocates fragmented leaves to large jobs.
Empirically, under Dynamic-MIG the \textit{Average External Fragmentation Delay} accounts for about 15\% of total makespan in large-dominant scenarios, dropping to 8\% in balanced and 4\% in small-dominant traces.
We also observe frequent reconfigurations in small-dominant and balanced traces (about {14} per run) versus only {5} in large-dominant traces.
Given that each scenario includes {62–64} jobs on two GPUs, Dynamic-MIG triggers reconfiguration quite often when small jobs are prevalent.

We further compare Flex-MIG and Dynamic-MIG across all training/inference mixes and workload-size distributions in Figure~\ref{fig:fm_dm}.
For these scenarios, we switch to {Aggressive Backfilling} to increase placement opportunities when reconfiguration is restricted by latency-sensitive inference jobs.
The overall trend persists: in 80–90\% of scenarios, Flex-MIG incurs only a 7–10\% increase in JCT relative to Dynamic-MIG, yet it consistently and substantially reduces waiting time.
Consequently, Flex-MIG shortens makespan by nearly 10\% in balanced and large-dominant groups and by about 20\% in small-dominant groups (Figure~\ref{fig:fm_dm_violin_makespan}), yielding higher throughput and utilization overall.

These results suggest that Flex-MIG’s one-to-many operation increases allocation flexibility through logical aggregation across MIG instances, which introduces a trade-off between slightly degraded per-job performance and reduced waiting time.
However, the impact of shorter queues dominates, ultimately improving cluster-level makespan and overall utilization.


\begin{figure}[h]
    \centering
    \includegraphics[width=\linewidth]{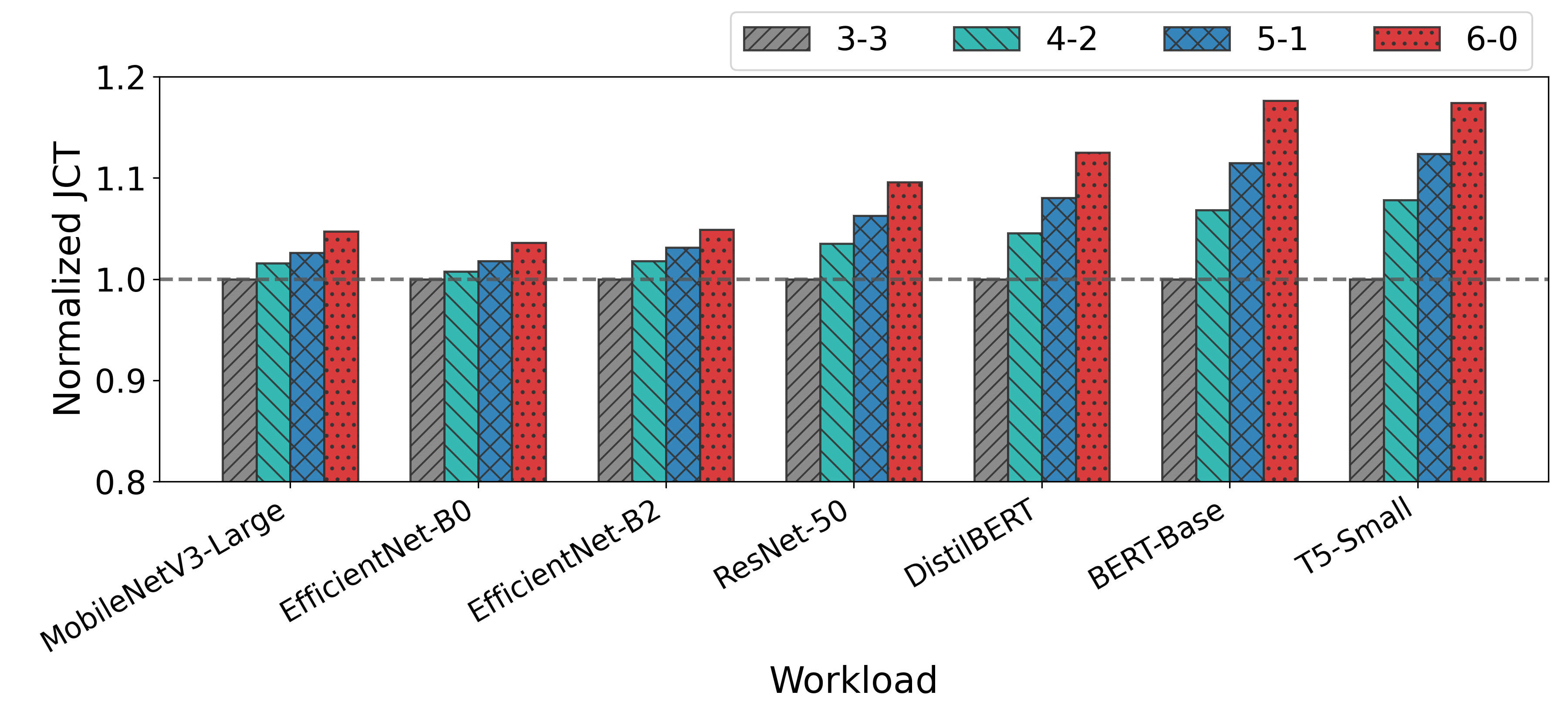}
    \caption{
    Normalized job completion time (JCT) for workloads of size~6 under different placement configurations. Each $N$–$M$ label indicates the number of MIG instances assigned to GPU~0 and GPU~1, respectively. 
}

    \label{fig:non-uni-placement}
\end{figure}  

\begin{figure}[h]
    \centering
    \includegraphics[width=1\linewidth]
    {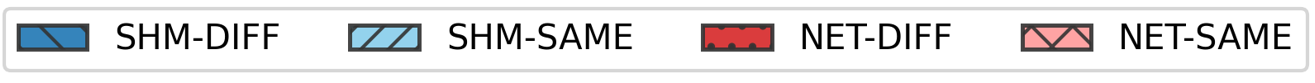}
    \begin{minipage}[t]{0.49\linewidth}
        \centering
        \includegraphics[width=\linewidth, height=3cm]{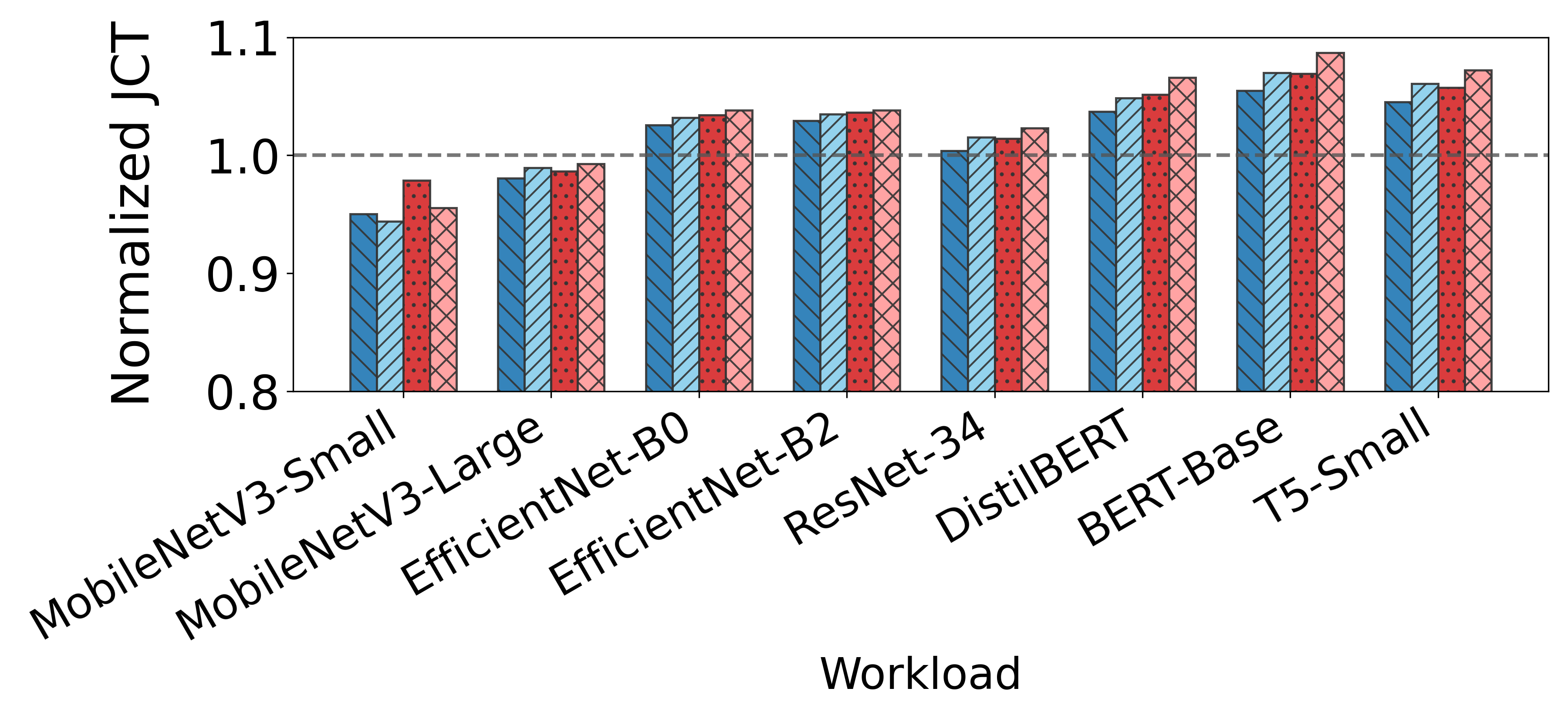}
        \subcaption{Single-job execution}
        \label{fig:single_jct_2}
    \end{minipage}
    \hfill
    \begin{minipage}[t]{0.49\linewidth}
        \centering
        \includegraphics[width=\linewidth, height=3cm]{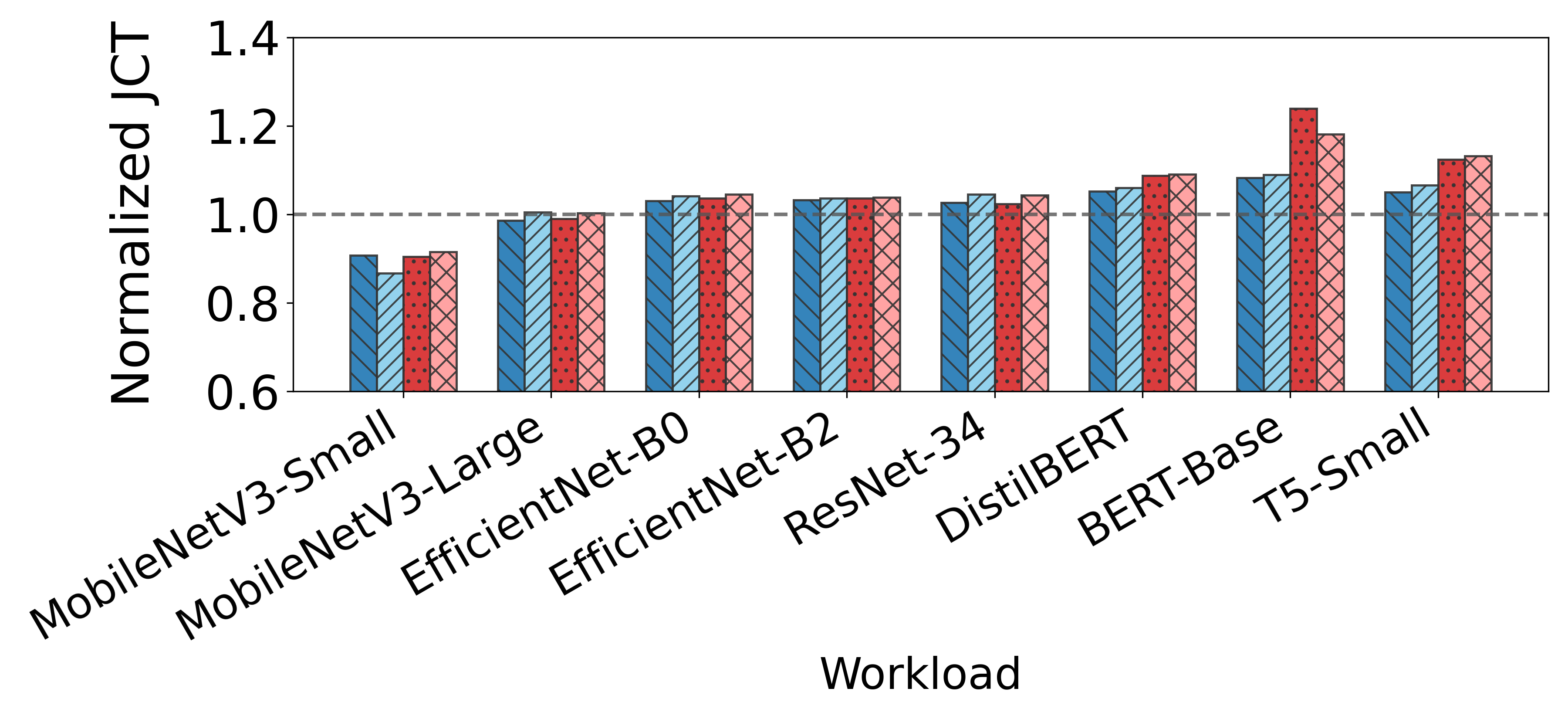}
        \subcaption{Full-capacity concurrent execution}
        \label{fig:multi_jct_2}
    \end{minipage}
    \caption{
    Comparison of one-to-one and one-to-many executions for workloads of size~2. 
Each bar represents a placement and transport combination: SHM (host shared memory) or NET (network-RDMA) communication, and SAME or DIFF placement depending on whether the two MIG instances reside on the same GPU or on different GPUs, respectively.
}
    \label{fig:jct_compare}
\end{figure}

\subsection{Job-level Analysis}\label{sec:job_perf}
This section presents the experimental evidence supporting the \textit{Topology-Aware Placement} policy introduced in Section~\ref{sec:orchestration-layer}. 
We then compare the performance differences between the one-to-one and one-to-many execution models and analyze how performance contention arises when multiple workloads execute concurrently.

\textbf{\textit{Impact of physical placement~(Topology)}}

Figure~\ref{fig:non-uni-placement} shows the performance of size-6 workloads when their MIG instances are placed differently across physical GPUs. 
As shown, performance degrades as instances become more unevenly concentrated on a single GPU. 
This degradation is attributed to bandwidth saturation at the per-GPU PCIe interface level, and the observation motivates the topology-aware placement rule incorporated into Section~\ref{sec:orchestration-layer}.

\textbf{\textit{One-to-one vs One-to-many Execution trade-off}}

Figure~\ref{fig:jct_compare} compares per-job performance between the conventional one-to-one model and our proposed one-to-many model in Flex-MIG. 
As model size increases, the communication cost per iteration grows, amplifying the performance gap between the two configurations (Figure~\ref{fig:single_jct_2}). However, the gap remains modest, typically below 10\%. In some workloads, network-based communication across different GPUs (NET-DIFF) even outperforms SHM-based communication within the same GPU (SHM-SAME).

When multiple workloads run concurrently (Figure~\ref{fig:multi_jct_2}), however, SHM consistently outperforms NET in all cases. 
This indicates that under concurrent execution, SHM communication experiences substantially less contention than network-based transport, resulting in higher overall efficiency for one-to-many execution in Flex-MIG.

\subsection{Runtime-level Microbenchmarks}\label{sec:micro_perf}
This section evaluates how much performance improvement SHM achieves over conventional network-based~(NET-RDMA) collective communication. 
Figure~\ref{fig:compare_net_shm} presents the measured bandwidth of the \texttt{AllReduce} operation when using 2 and 8 MIG instances. 
In both cases, SHM-based transport achieves the highest bandwidth when MIG instances reside on the same physical GPU, outperforming NET-based communication across GPUs. 
Similar results are observed with 4 and 6 instances, confirming that SHM consistently provides superior performance for intra-GPU collectives. 
This improvement validates that our NCCL modification effectively unlocks high-bandwidth, low-latency communication paths among MIG instances, which plays a key role in enabling efficient distributed execution in Flex-MIG.
\begin{figure}[h]
\centering
\begin{subfigure}[t]{0.48\linewidth}
    \includegraphics[width=\linewidth, height=3cm]{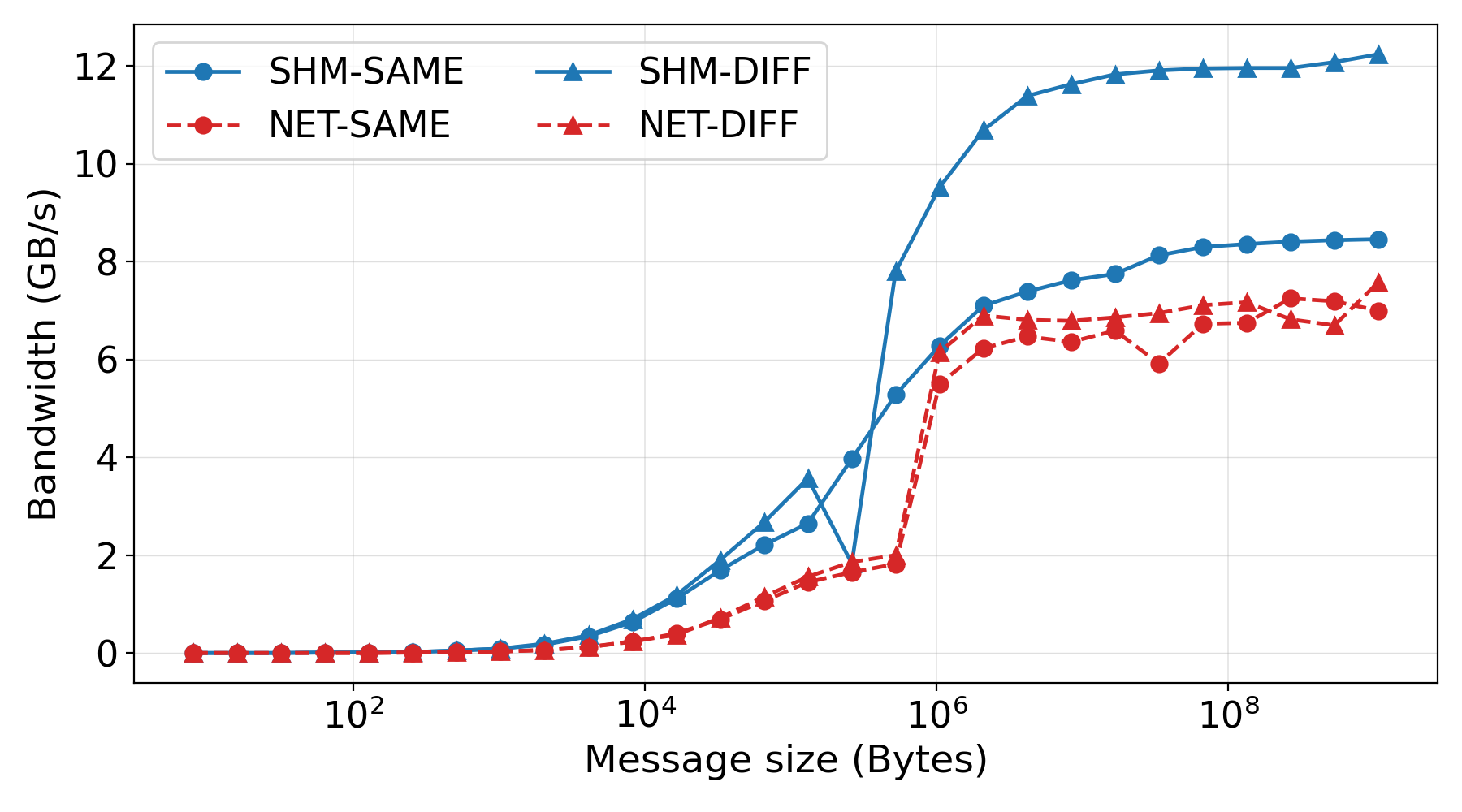}
    \subcaption{2 MIG instances}    
\end{subfigure}
\hfill
\begin{subfigure}[t]{0.48\linewidth}
    \includegraphics[width=\linewidth, height=3cm]{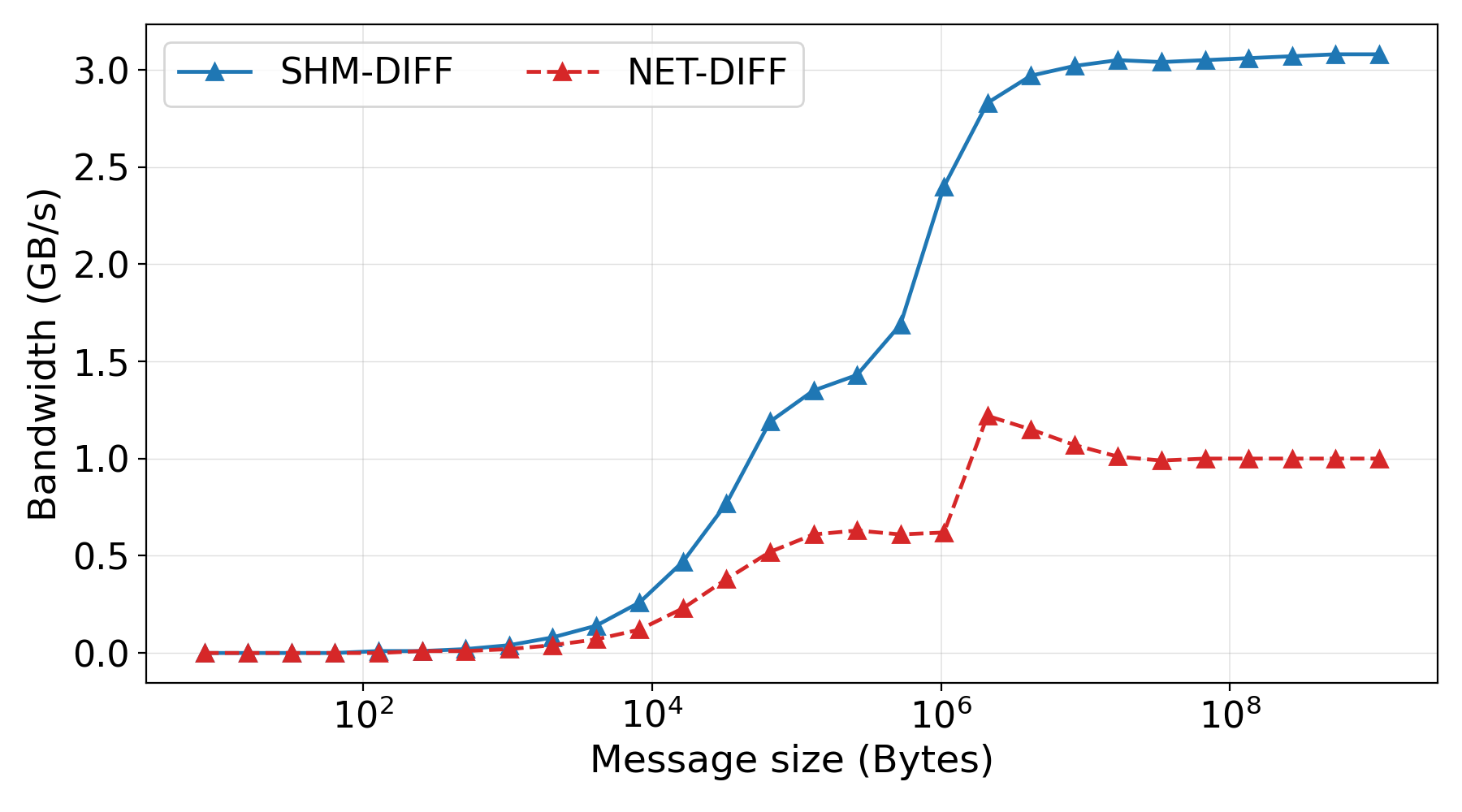}
    \subcaption{8 MIG instances}    
\end{subfigure}
\caption{Bandwidth comparison between SHM and NET~(RDMA) transports across collective operations and MIG configurations.}
\vspace{-1em}
\label{fig:compare_net_shm}
\end{figure}

%% file: sections/7_Relatedwork.tex
\section{Related Work}\label{sec:related_works}

\textbf{\textit{Software-level sharing.}}
Prior systems such as MPS\cite{nvidia-mps}, TGS\cite{TGS}, and Orion\cite{Orion} perform fine-grained multiplexing of GPU resources to improve utilization in multi-tenant settings.
While effective in increasing concurrency, these approaches rely on shared caches and memory bandwidth, resulting in inter-job interference and limited isolation.

\textbf{\textit{Dynamic composition atop MIG.}}
NVIDIA MIG~\cite{nvidia-mig} provides hardware-level isolation, yet its rigid predefined profiles and drain-required reconfiguration limit flexibility in multi-tenant clusters.
To further overcome MIG’s lack of elasticity, recent systems combine MIG with MPS to recover unused capacity: MIGER~\cite{MIGER} jointly decides MIG partition sizes, job co-location, and per-job MPS shares (i.e., many-to-one sharing within a MIG instance), whereas ParvaGPU~\cite{parva} explicitly enables MPS within each MIG instance and scales the number of identical processes to maximize intra-instance efficiency.

\textbf{\textit{Our position.}}
Unlike these approaches, Flex-MIG redefines the MIG operational model: it executes a single job across multiple fixed leaves, thereby avoiding drain-required reconfiguration while maintaining hardware isolation. This turns MIG from a static partitioner into a composable runtime resource layer.

%% file: sections/8_Conclusion.tex
\section{Conclusion}
We revisited the fundamental limitations of MIG’s one-to-one allocation model and showed that fixed profiles, tree-constrained merging, and drain-required reconfiguration collectively degrade cluster-wide utilization.
\textit{Flex-MIG} introduces a software-only one-to-many model that fixes GPUs to minimal leaves and aggregates multiple instances per job.
By flattening all MIG resources within a host, Flex-MIG enables logical aggregation across instances, resolving resource fragmentation and eliminating the need for physical reconfiguration.
It further extends NCCL to support shared-memory collectives among MIG instances, preserving isolation while enabling efficient intra-GPU communication.
Our evaluation—validated against real measurements—shows that Flex-MIG improves makespan by up to 17\% over Dynamic- and Static-MIG baselines, with only modest per-job overhead.
These results highlight that rethinking MIG as a logically composable, software-managed resource layer can unlock substantial efficiency gains in multi-tenant GPU clusters.